\documentclass[12pt,preprint]{aastex}

\shorttitle{Variability of Narrow-Line Seyfert 1s}
\shortauthors{Klimek, Gaskell, \& Hedrick}

\begin{document}

\title{OPTICAL VARIABILITY OF NARROW-LINE SEYFERT 1 GALAXIES}

\author{ELIZABETH S. KLIMEK, C. MARTIN GASKELL \& CECELIA H. HEDRICK}
\affil{Department of Physics \& Astronomy, University of Nebraska,
Lincoln, NE 68588-0111} \email{penumbra89@hotmail.com,
mgaskell1@unl.edu, piqueen314@hotmail.com}

\begin{abstract}

We present results of a broad-band photometric study of the
optical variability of six Narrow-Line Seyfert 1 (NLS1) galaxies
observed at 172 epochs.  We searched for microvariability on 33
nights. Strong evidence for microvariability was found only for
our lowest luminosity object, NGC 4051, on one night.  Weaker
evidence suggests such variability on a few other nights for two
other objects, but the data are not as convincing.  Intra-night
variability in NLS1s is thus concluded to be rare and of low
amplitude.  We give illustrations of how variable image quality
can produce spurious variability.  We find that for well-studied
non-NLS1s there is a spread in the amplitude of seasonal
variability (i.e., in some years an AGN is more variable than in
others). We find that the means of the variability amplitudes of
non-NLS1s over several seasons vary from object to object (i.e.,
some AGNs are, on average, more variable than others). NLS1s also
show a spread in seasonal variabilities. The best-studied NLS1,
Ark 564, shows a range of amplitudes of variability from season to
season that is comparable to the range found in BLS1s, and in one
season Ark 564 was as variable as the most variable non-NLS1. The
seasonal amplitudes of variability for NLS1s are mostly in the
lower half of the range of non-BLS1 seasonal amplitudes, but the
absence of a suitable control sample makes a precise comparison
difficult. However, on long timescales (weeks to years) NLS1s as a
class are not {\it more} variable than non-NLS1s. The extreme
variability seen in the X-rays was not seen in the optical.  This
has consequences for the models of AGNs in general as well as
NLS1s in particular.

\end{abstract}


\section{INTRODUCTION}

It has long been realized that some AGNs only have narrow emission
lines in their spectra while at the same time showing the
characteristic spectrum of the broad-line region (e.g., Phillips
1977, Koski 1978, Davidson \& Kinman 1978).  Thus the BLR (high
density) lines were narrower than usual.  These objects came to be
called Narrow Line Seyfert 1s (NLS1s) (Gaskell, 1984; Osterbrock
\& Pogge, 1985, Goodrich 1989).  Although there is no sharp
demarcation, an AGN is commonly call a NLS1 if the broad lines
have a FHWM $\le 2000$ km~s$^{-1}$. We will refer to AGNs with
FWHM $> 2000$ km~s$^{-1}$ as BLS1s (``broad-line Seyfert 1s'').
The properties of NLS1s lie at one end of a set of correlations
between AGN properties that is commonly called eigenvector 1
(Boroson and Green, 1992). In addition to the curiously narrow
``broad'' lines, NLS1s tend to have steeper soft X-ray spectra
than non-NLS1s (Boller, Brandt \& Fink 1996). Many exhibit rapid
soft X-ray variability, which can also be large in amplitude, such
as a factor of $\sim 100$ in a day in IRAS 13224-3809 (Boller et
al. 1997). NLS1s show strong optical Fe II emission lines (Sargent
1968) and sometimes the higher ionization Fe lines (Davidson \&
Kinman 1978), but while the Fe II equivalent widths are about the
same as those of non-NLS1s (Gaskell 1985), the H$\beta$ equivalent
widths are smaller than those of non-NLS1s (Gaskell 1985, Goodrich
1989). The [OIII]/H$\beta$ line ratio is $< 3$, which is less than
the dividing line Shuder \& Osterbrock (1981) found between
Seyfert 1s and Seyfert 2s (i.e., the [OIII]/H$\beta$ ratio implies
that they are Seyfert 1s).

A common trait possessed by all AGNs is that that they display
some degree of variability (Ulrich, Maraschi, \& Urry 1997;
Gaskell \& Klimek 2003). The most vigorous variability observed is
that of the X-rays (Mushotsky, Done, \& Pounds 1993). Variability
is valuable in that it provides clues to understanding what AGNs
actually are and how they work. Details of the inner workings of
AGNs remain poorly understood, despite over three decades of
research. Variability can help set constraints on the sizes of
different regions of AGNs and can give information about the
processes that are driving the variations. There has been much
effort in looking for optical variability, some of it in
conjunction with monitoring in other wavebands.  If a link is
found between the variability in different wavebands, then the
processes behind each kind of variability are probably related.
Explanations of AGN variability include hotspots on accretion
disks, flares, and relativistic jets.  BL Lac objects and
optically violently variable AGNs (OVVs) exhibit strong
variability on short timescales.  This is believed to be a
consequence of relativistic beaming by jets.  The beaming can
amplify intrinsic variations, which may or may not originate from
within the jet.

Rapid (sub-diurnal) variability in the X-ray emission of AGNs is
well-known, but there have been conflicting claims about the
frequency of occurrence of rapid intra-night optical variability.
So far, observing campaigns searching for optical variability in
AGNs have typically yielded low amplitude variations on timescales
of no less than a few days (e.g., Webb \& Malkan 2000a).  De
Ruiter \& Lub (1986) did not find any rapid ($< 1$ day) variations
of greater than 0.5 \% for any of the eight Seyfert galaxies they
observed, while other observers have reported low-level
microvariability on sub-diurnal timescales.  For example,
Merkulova (2000) reported sub-diurnal variability at the 1\% level
on 60\% of nights for NGC 4151.  Jang \& Miller (1995, 1997) found
that 8 out of 17 radio-quiet and 6 out of 7 radio-loud AGNs showed
intra-night variability on the order of a few percent (see also
Carini, Noble, \& Miller, 2003).  Merkulova  (2000) concluded that
intra-night variability is transient in character and has
manifested itself with different probabilities for different
galaxies.  Even though there is some uncertainty over the
frequency of occurrence of sub-diurnal variability, it would seem
that extremely rapid and/or large amplitude optical variations are
rare.

The most extreme X-ray variability in non-OVV AGNs is seen in
NLS1s.  The shortest timescales for X-ray variability are about
200 to 1000 s (Boller et al. 1993).  Over a short period of time,
the amplitude can change by as much as a factor of 4, as seen in
the NLS1 IRAS 13224-3809, which varied by this much over a course
of hours (Boller et al. 1997).  Large amplitude rapid X-ray
variability is one of the interesting properties of NLS1s.  It has
been suggested that NLS1s are beamed (Boller et al. 1997), since
they display similarly strong variations in the X-rays.

If rapid optical variations are present in AGNs other than BL
Lacs, it is possible that they would most likely be found in
objects displaying the most extreme X-ray variability, namely the
NLS1s.  On long timescales Giannuzzo \& Stirpe (1996) and
Giannuzzo et al. (1997) compared the {\it Balmer-line} variability
of NLS1s with that of NGC 5548 and found the NLS1s to be less
variable. However, this was not a general comparison with
non-NLS1s, and it was possible that NGC 5548 was more variable
than the average BLS1. Young et al. (1999) unsuccessfully searched
for intra-night optical variability in IRAS 13224-3809, while
Miller et al. (2000) reported dramatic variability in the same
object on one night.

The work presented here has two goals: (a) to look for evidence of
sub-diurnal optical continuum variations, especially those of
large amplitude, in a larger sample of NLS1s and (b) to see
whether NLS1s are more variable in the optical than non-NLS1s on
longer timescales. For one of the objects in our sample, Ark 564,
there are observations over a much longer period of time, so we
discuss this object at more length in a separate paper (Gaskell et
al. 2004).

\section{OBSERVATIONS}

The majority of the observations were made with the University of
Nebraska's  0.4-m Schmidt-Cassegrain telescope of the university's
Lincoln Observatory in a bright sky location.  Images were
recorded with a Kodak KAF-0401 CCD, giving 0.887 arcseconds per
pixel through a f/5 focal reducer.  All images were taken through
a standard Johnson V filter.  Additional observations were made
with the 0.8-m Cassegrain at the university's Behlen Observatory
near Mead, Nebraska, in a darker sky location, with a Kodak
KAF-1001E chip giving 0.59 arcseconds per pixel with an f/9 focal
reducer through an identical filter.  The NLS1s and the comparison
stars were measured through photometric apertures of identical
effective solid angles and we found that no scaling was necessary
between observations from the two telescopes.

A sample of six NLS1s was chosen based only on the criteria that
they had to be bright enough to be observed with the 0.4-m
telescope and that they had to be high declination sources in
order to maximize the amount of time that they would be accessible
during the night. Thus, our selection was not based on any other
specific property or characteristic of this class of AGNs. In
particular, variability history was not taken into consideration.
The objects selected were Ark 564, Mrk 478, Mrk 493, Mrk 335, Mrk
359, and NGC 4051.

Observations for most objects were taken between May 2000 and June
2003, with the exception of Ark 564, for which the observations
considered here began in August 1998.  In searching for
intra-night variations, we observed objects continuously for most
of a night.  Integration times were 10 to 15 minutes on the 0.4-m
telescope and 5 minutes on the 0.8-m telescope.  Anywhere from
three to 26 images per object were obtained during any one night,
giving a range of 30 minutes to 6.5 hours of continuous data for
the most intensively observed objects.

\section{DATA REDUCTION}

\subsection{Comparison Stars}

A summary of the adopted magnitudes for the comparison stars used
for the NLS1s (except Ark 564 –- see Gaskell et al. 2004) is given
in Table 1, and finding charts are shown in Figures 1 -- 5. Where
possible, comparison stars were chosen from the literature. For
Mrk 359, Mrk 493, \& Mrk 478 we calibrated our own comparison
stars from the count rates on a handful of our best nights.  The
relative calibration of these stars is good, but the absolute
calibration is believed to be uncertain by ±0.2 magnitudes.  This
uncertainty has no impact on this study, as we are only interested
in changes in magnitudes.  In the case of Mrk 478, our magnitudes
agree with the relative magnitudes given by Webb and Malkan
(2000b).

The detector noise is non-Gaussian because of hot pixels and
cosmic rays.  If one comparison star gave anomalous magnitudes for
an image, it was not used for that image.  Likewise, if the
average magnitude for an AGN was significantly off for an image on
a night, it was dropped from the nightly average.  Between 5 \%
and 10 \% of comparisons were discarded.  Visual inspection often
revealed obvious image problems in these cases.

\subsection{Size of Photometric Aperture}

In order to maximize the accuracy with which the counts from an
object are measured with respect to a sky background, the size of
the measuring aperture should be small in order to reduce the sky
background contribution, but at the same time, large enough to get
a good signal-to-noise ratio and to minimize the effect of
fluctuations in the measurement due to miscentering.  A compromise
must then be chosen in order to obtain the best measurement.

For point-sources in stellar photometry, these effects have been
studied by Howell (1989).  For faint stellar sources, the optimum
aperture radius is about three pixels or about the FWHM.  A number
of previous studies of AGN variability have followed the Howell
prescription (e.g., Carini et al, 1991; Jang \& Miller, 1995,
1997), but, as Cellone, Romero, \& Combi (2000) pointed out, the
effects of the underlying host galaxy, especially in low
luminosity AGNs, should not be ignored.  Because of the underlying
host galaxy even small seeing fluctuations can introduce spurious
variability in the AGN flux that can be mistaken for
microvariability.  This is because poor seeing turns a point image
into a extended image, but a galaxy already has an extended image
and is therefore affected much less.  This means that poor image
quality causes more light loss from a circular aperture for a star
than for a galaxy, and thus the galaxy will appear to be brighter
relative to the star.  We provide illustrations of this effect
below. Similar conclusions in the spectroscopic case of choosing a
suitable aperture were found by Peterson et al (1995).

Since the ideal aperture is a function of the radial brightness
distribution of the galaxy and its brightness relative to the AGN,
we experimented with two aperture sizes to find the one most
appropriate for a given object.  The larger aperture had a radius
of 9 pixels on the 0.4-m telescope, corresponding to 8-arcseconds,
and the smaller aperture had a radius of 5 pixels, or
4.4-arcseconds.  The same sky annulus was kept in both cases,
originally chosen to be appropriate for the larger aperture size,
in order to exclude most of the galaxy component from the sky
background measurement.  The annulus had an inner radius of 13.5
pixels (12-arcseconds) and an outer radius of up to 22.5 pixels
(20-arcseconds).

Each object was measured with both apertures and the resulting
errors were analyzed.  We only give here the measurements
resulting from the aperture that produced the lowest rms variation
in the errors of the nightly means for each object.  For all
objects but Mrk 478 (one of our faintest and most compact objects)
the larger aperture of 8-arcseconds was used.  A summary of our
individual observations on nights we searched for microvariability
is presented in Tables 2 -- 6. We discuss the estimation of our
errors below. For other nights just the nightly means are given in
Table 7.

\section{ANALYSIS}

There are several sources that can give rise to spurious
variability.  These sources include imperfect flat-fielding,
possible inconstancy of comparison stars, and changes in image
quality due to seeing and focus shift.  The latter causes
fluctuation in the galaxy component measured (see above).

\subsection{Estimation of Errors}

Photon statistics alone underestimate errors so we used other
methods to estimate the errors.  For the long-term light curves,
the error for each night was calculated by dividing the standard
deviation ($\sigma$) of the magnitudes by the square root of the
number of images (n) for the night.  Nights with few images
typically have the greater error, but sometimes such nights had
fortuitously small errors (e.g., if two or three of the magnitudes
happened to be essentially the same).  In these cases the
individual night's $\sigma$ was replaced by the median standard
deviation $\sigma_{med}$ for all of an object's nights of
observation. The $\sigma_{med}$ was thus used as a more likely
estimator for calculating each of these night's errors in order to
prevent a serious underestimation of the errors.

This method of estimating errors obviously cannot be used when
searching for intra-night variability, since it calculates the
errors under the assumption of no variability.  Looking at the
differential light curves of stars comparable in brightness to the
AGN is a useful and important check but gives only a lower limit
to the error because the additional error due to the host galaxy
(see above) is not taken into consideration.

In order to estimate the errors on nights that were searched for
intra-night variability, the following method was used.  The
difference of adjacent magnitudes in the time series, $\Delta$$m_i
= m_i - m_{i-1}$, was used to find the point-to-point variation
during the night.  The standard deviation of this difference was
then divided by the square root of two, since a difference of two
magnitudes was taken, to give the estimated error.

If there is no variability during the course of the night, then
this error will be indistinguishable from the nightly standard
deviation.  This was in fact the case for all of our intra-night
variability search nights, with the exception of the one night
where intra-night variations were detected for NGC 4051.

The difference method could give erroneously large errors if any
rapid variations are present on a microvariable timescale of
$10 –- 15$ minutes. However, we consider this to be unlikely (see
below).

\subsection{Searching For Variability}

While it is difficult to rule out microvariability, it is easier
to detect or rule out variability on longer timescales (hours).
AGNs are already known to vary on timescales of more than a day.
Therefore, it can be reasoned that if AGNs show variability on a
smaller timescale of minutes, then they will show variability on
the intermediate timescale of hours.

To search for variability on the timescale of a few hours, two
statistical tests were performed.  First, the magnitudes for a
night were divided into first and second halves.  A Student t-test
was used to evaluate the significance of differences in the mean
magnitude between the two equal halves of the night.  The second
test was to evaluate the significance of the correlation
coefficient for the magnitudes versus time.  A significant
correlation would mean a significant possibility that variations
occurred during a night.  If either test gave a two-tailed
significance of $> 90$\% (i.e., a less than 10\% probability of
arising by chance), further investigation was conducted.

The t-test and the correlation test are only sensitive to a
general trend during the night.  Other types of variation would be
missed (e.g., a rise and fall or sinusoidal oscillations).  To
search for general variations on a timescale of $30 -— 45$
minutes, an F-test was conducted to compare the variance of the
magnitudes during a night with the variance in their
point-to-point differences.  This tells whether or not there is
significant correlated variability between exposures, which
correspond to $10 - 15$ minute intervals.  Again, we conducted
further investigation if the F-test gave a significance of $>
90$\%.

\subsection{Systematic Errors}

In the cases where the above-mentioned statistical tests showed a
high level of significance, we investigated the effects of
possible sources of systematic error.

\subsubsection{Image-Quality Effects}

As is well-known, good flat fielding is difficult to attain, even
when using sky flats.  Variations of 1\% or more across the CCD
chip were not uncommon.  However, this effect was minimized by
having several (typically four) comparison stars, by trying to
choose comparison stars that were close to the AGN, and by
positioning the AGN on the same part of the chip every night as
best as possible.  Images that were found to have significant
flat-fielding problems were removed from our analysis. This
sometimes involved dropping entire nights in order to prevent
false detections of variations.  Where there were signs of
possible microvariations, possible flat-fielding problems were
investigated by looking at the (x,y) position of the AGN on the
chip in addition to visual inspection. If the object ``wandered''
due to inconsistent positioning during a night, its position was
plotted versus time to see if its behavior correlated at all with
the object's light curve.

The effect of changes in the image quality due to seeing or focus
changes (see section 3) were examined by comparing the trend of
the average FWHM of the comparison stars during the night with the
object's light curve over the same period of time.  For each night
the AGN magnitudes were also plotted against seeing to further
check for any correlation.  Such a correlation would signify that
any variations present were spurious.  Two examples of this
analysis are given below.

Only two of the objects, Mrk 359 and Mrk 478, showed a clear FWHM
dependence of the magnitudes.  This dependence was only present in
the 4.4-arcsecond aperture.  No effect was seen in the 8-arcsecond
aperture.  Figs. 6 -- 10 show two examples of the seeing
dependence during a night of Mrk 359 and Mrk 478 observations.
Figs. 6 and 9 are plots of the light curves obtained by using a
4.4-arcsecond aperture.  Fig. 7 shows the average FWHM of the
comparison stars from image to image for Mrk 359. Figs. 8 and 10
confirm the correlations between the magnitudes of the AGNs and
the FWHMs.  It can be seen that if the changes in image quality
were not taken into account, there would be a spurious detection
of variability.  The statistical tests showed no significant
variation for the large apertures, while for Mrk 359, for example,
the Student t-test was highly significant for the smaller aperture
(98\% significance).

Analysis of the errors for all the nights of observations for Mrk
359 showed the least sub-diurnal variability for the larger
aperture, and this gave the smoothest long-term light curve.
However, in the case of Mrk 478, the smaller aperture gave the
most consistent errors between nights.

\subsubsection{Constancy of Comparison Stars}

The constancy of the comparison stars was checked by plotting the
magnitudes of each individual star with respect to the averages of
the other comparison stars for each night.  For NGC 4051 Star 5
showed evidence of a slow change of 0.03 magnitudes over three
observing seasons and was therefore not used.  For NGC 4051, for
which the only instance of intra-night variability was found (see
below), it was especially important to have dependably constant
stars. There was also slight possible variations for two
comparison stars of Ark 564 (see Gaskell et al. 2004) but these
are comparable to our measuring errors and have a negligible
effect on our results, since for Ark 564 we are averaging
comparisons with four stars.

\section{RESULTS}

\subsection{Sub-Diurnal Timescales}

All nights containing at least a dozen images of the same object,
amounting to at least three hours' worth of continuous monitoring,
were searched for evidence of microvariability.  The significance
was evaluated as described in 4.2. The results are shown in Table
8.  For each test we give the probability of the possible
variability arising by chance (i.e., the null hypothesis is no
variability).  As can be seen, the F-test reveals no correlated
variation on the time scale of $\sim 30$ minutes.  The Pearson
correlation coefficient test and the Student t-test reveal trends
with one-tailed significances of $> 90$\% on a time scale of
several hours on five nights.  However, because we are looking at
33 nights of monitoring, we would expect to get a significance of
1 in 10 about three times.

Allowing for the number of nights of observation, it is apparent
from Table 8 that there is only one night with a high probability
of microvariability: NGC 4051 on 2003 Feb 20.  There are four
other nights where the correlation and t-tests indicate possible
variability at the $\sim 90 –- 95$\% significance level.  Apparent
variations on these nights were checked for spurious causes.

A 0.045 magnitude change was found in the light curve of NGC 4051
during the course of one night (see Fig. 11).  Both the
correlation and t-tests gave greater than 99.6\% confidence that
the variations were not due to chance.  We checked for possible
instrumental causes.  Fig. 11 shows the behavior of NGC 4051 on
2003 February 20 with respect to star 2, the brightest comparison
star, and with respect to the average of the other three
comparison stars.  Both curves are in good agreement, attesting to
the reliability of the comparison stars.  The larger-than-average
fluctuation in the third image is caused by a problem with star 2.
Such problems are not unusual (see 3.1) and are consistent with
our detector noise characteristics.  Likewise, statistically there
could also be a problem with the AGN in one or both of images 8 or
9.

Figures 12 and 13 show plots of the y-position on the chip and the
FWHM as a function of time.  Although there are slight changes in
the y-position during the night, comparable changes in the
position on the chip are seen on other nights without affecting
the magnitude.  The two largest changes in the light curve (Fig.
11) and the y-position (Fig. 12) on the chip occur simultaneously.
However, after this point the two plots behave differently.  There
were no peculiarities found in either the flat-fielding or in the
dark subtraction.  The x-position change (not shown) also shows no
detailed correlation with the light curve.  The image quality
improved during the night, but the FWHM does not correlate in
detail with the shape of the light curve. The 8-arcsecond radius
aperture magnitudes show no correlation with FWHM on any other
night for NGC 4051 (or indeed for any other object).  When the
4.4-arcsecond radius aperture magnitudes do show a correlation
with FWHM (see Figs. 8 and 10), it is in the {\it opposite} sense
to what was found here.  We are thus unable to find artificial
explanations of the apparent sub-diurnal variations of NGC 4051 on
2003 February 20.

With the exception of the one night for NGC 4051 we therefore find
very little evidence for variability on an intra-night timescale.
A study of microvariability in Seyferts (Carini, Noble, and
Miller, 2003) found only one Seyfert, the BLS1 Ark 120, in a
sample of eight, which showed signs of sub-diurnal activity. Their
result for the NLS1 Mrk 335 was consistent with the null result
obtained from our study.  Jang and Miller (1997) also found no
evidence of microvariability in Mrk 335 in their study of
radio-quiet versus radio-loud quasars.  Webb \& Malkan (2000a)
searched for intra-night variability in Seyfert 1s.  No evidence
for such variability was found, but they typically had only two
images per night, compared with the more than twelve in this
study, and their observational errors were about ± 0.03
magnitudes.  Their sample included Mrk 478, an object for which we
found no microvariations.

\subsection{Longer Timescales}

Long-term light curves from the nightly mean magnitudes given in
Table 7 are plotted in Figs. 14 -- 18.  As can be seen,
variability is present for all six objects on a timescale of
months to years.

In all photometric and spectroscopic studies of continuum
variability there is contamination from the host galaxy.  This
varies with the size of the photometric or spectroscopic aperture
used.  In order to make a legitimate comparison, our magnitudes
were scaled to small aperture sizes, which have been used in most
spectrophotometric studies such as those of the {\it International
AGN Watch}. This was done by taking some of our best images for
each object (i.e., images having the lowest FWHM) and measuring
the counts for successively smaller apertures.   The ratio between
the original aperture we used and the smallest practical aperture,
2.7 arcseconds, was found.  This ratio was then used to scale our
data by removing the extra host galaxy contribution in the larger
aperture.  This effectively gives a multiplicative correction
factor for the amplitude of variability. These scale factors are
given in Table 9.  In Table 10 we give scaled seasonal standard
deviations for our six NLS1s and also seasonal standard deviations
for NGC 4051 from the International AGN Watch observations of
Peterson et al. (2000).  In Table 11 we give the seasonal standard
deviations for eight non-NLS1 AGNs observed by the International
AGN Watch and others.

A comparison of the seasonal standard deviations of the NLS1s with
the sample of BLS1s is given in Fig. 19.  From this we note the
following:

\begin{enumerate}

\item For well-studied BLS1s (e.g., NGC 5548, NGC 4151, 3C 273),
there is a range in standard deviation from season to season.

\item Some BLS1s appear on average to have higher seasonal
standard deviations than others.

\item NLS1s also show a spread in seasonal standard deviations.

\item The NLS1 seasonal standard deviations are in the lower-half
of the range of BLS1 standard deviations.

\item The well-studied NLS1 Ark 564 shows a range of seasonal
standard deviations comparable to that of a BLS1 and in one season
was as variable as the most variable BLS1 (see Gaskell et al. 2004
for more detailed discussion).

\end{enumerate}

\section{DISCUSSION}

\subsection{Short-Timescale Variability}

Our study clearly shows that high-amplitude variability in NLS1s
is rare.  As noted already, Miller et al. (2000) did however find
an example of large-amplitude, rapid optical variability in IRAS
13224-3809. With this exception, NLS1s do not display rapid
high-amplitude variability in the optical such as is seen in BL
Lac objects or in the X-rays of NLS1s.  Such short timescale
variability as there is must usually be of low amplitude.  The
study of Ferrara et al. (2001) also supports this conclusion.
Interestingly, the one NLS1 in which we find reasonable convincing
evidence for intra-night variability is also the least luminous
AGN in our sample.  It is therefore the one with the least massive
black hole (Peterson et al. 2000) and the one for which the most
rapid variability would be expected. It is also perhaps
interesting that this microvariability in NGC 4051 occurred when
there was a rapid (1-day) drop in the flux (see Fig. 14).

Factors contributing to spurious artificial variations are
numerous, and great care must be taken to account for them.  The
inhomogeneity of data in previous variability studies is a
problem.  The errors must be well-determined, and all sources of
systematic error must not be ignored.  It is important to reduce
errors because true variations could be hiding in the noise.
Convincingly detecting microvariations is no small task.

Because of uncertainties in studies of non-NLS1 microvariability,
it is not yet possible to make as detailed a comparison of the
microvariability of NLS1s and BLS1s as we would like.  As usual,
more data are needed to further strengthen our findings.  With all
the factors that can lead to a false positive detection of
microvariability, it is especially beneficial to have multiple
observers to verify a detection of microvariability.  This has
generally not happened in the past.

\subsection{Long-Term Variability}

From Fig. 19 it might seem that, on average, NLS1s are less
variable than BLS1s, but before such a conclusion can be made,
selection effects need to be understood.  None of our objects were
chosen on the basis of optical variability characteristics, but
BLS1s are typically chosen for studies with a hope that they will
show high-amplitude variations.  This selection is based on prior
variability history, and, as can be seen from Fig. 19, some
objects tend to be more variable than others.  Thus, there is a
selection effect if the non-NLS1 sample consists primarily of the
most variable BLS1s, in which case the comparison between the
variabilities of the two classes is not a valid one. Giannuzzo \&
Stirpe (1996) and Giannuzzo et al. (1998) compared the Balmer-line
variability of NLS1s with that of NGC 5548 and Fig. 19 shows that
NGC 5548 is indeed more variable than the average BLS1.

We believe that, because of the lack of an adequate control sample
of BLS1s, it is not yet possible to conclude that NLS1s are less
variable in the optical than BLS1s.  There is no evidence that the
amplitude of variability of NLS1s and BLS1s is different.

\subsection{Implications}

That NLS1s vary in amplitude by so much so rapidly in the X-rays
without any similar behavior in the lower energy end of the
spectrum is a remarkable result and says something important about
the structure of AGNs.  The fact that NSL1s probably do not vary
any differently than non-NLS1s in the visible continuum region is
in itself another interesting finding.

These two findings rule out a situation where the X-rays and the
optical variations have a simple common origin.  The source
producing the X-ray variability cannot be the same as that
producing the optical variations.  If they both come from an
accretion disk, for example, then they must be in distinctly
different parts.  If orientation plays a role in the NLS1
phenomenon, as it does through relativistic beaming, then only the
X-ray emission is being enhanced.

Our results also do not support simple reprocessing, since the
optical band should respond in some manner to the X-ray changes.
That is, if the X-rays vary with a larger amplitude in NLS1s than
in BLS1s, then the optical variations, if present, should also
have a larger amplitude in NLS1s, even if the increase is small.
We know that in general NLS1s are highly variable in the X-rays,
constantly varying rapidly, (this is certainly true for one of our
objects, Ark 564, Turner et al. 2001, Shemmer et al. 2001, Edelson
et al. 2003) but we do not see any significant rapid optical
variations in NLS1s (with the exception of one night for one
object).

\section{CONCLUSIONS}

Our data show that, as a class, there is no evidence that NLS1s
behave any differently than non-NLS1s in terms of variability.
NLS1s can exhibit signs of variability on a timescale of hours,
but such events are rare and of low amplitude.  They do not
exhibit the sort of remarkable variability seen in the X-rays.
Longer term variability over a timescale of days to months is the
norm, as is with non-NLS1s.

\acknowledgments

Financial support for this work was provided in part by the
University of Nebraska Layman Fund, the University of Nebraska
Undergraduate Creative Activities and Research Experiences
program, and by the National Science Foundation through grant AST
03-07912.  We are grateful to Beccie Grove and Matt Poulsen for
assistance in obtaining some of the observations.

\clearpage
\begin{deluxetable}{lrrlrlrrrrrr}
\tablewidth{0pt} \tabletypesize{\small} \tablecaption{Comparison
Stars\label{tbl 1-A_a}} \tablehead{ \colhead{Object} &
\colhead{Star} & \colhead{V (magnitude)}  & \colhead{Reference}}
\startdata
    &       &       &       \\
Ark 564 &   0   &   12.175  &   Gaskell et al. (2004)    \\
    &   1   &   13.658  &       \\
    &   3   &   14.176  &       \\
    &   5   &   14.490   &       \\
    &       &       &       \\
Mrk 478 &   2   &   14.255  &   uncertain to +/- 0.2    \\
    &   3   &   13.495  &       \\
    &       &       &       \\
Mrk 493 &   2   &   13.107  &   uncertain to +/- 0.2    \\
    &   4   &   13.684  &       \\
    &   5   &   13.933  &       \\
    &   8   &   14.429  &       \\
    &       &       &       \\
Mrk 335 &   4   &   13.679  &   Shrader et al. (1990)  \\
    &   6   &   13.794  &       \\
    &       &       &       \\
Mrk 359 &   1   &   13.473  &   uncertain to +/- 0.2    \\
    &   3   &   13.533  &       \\
    &   4   &   13.725  &       \\
    &   5   &   14.540  &       \\
    &       &       &       \\
NGC 4051    &   1   &   14.010   &   Penston, Penston, \& Sandage 1971    \\
    &   2   &   11.120   &       \\
    &   4   &   13.390   &       \\
    &   5   &   14.760  &        our calibration (not used)  \\

\enddata
\end{deluxetable}

\newpage
\begin{deluxetable}{lrrrrc}
\tablewidth{0pt}
\tablecaption{V-Magnitudes During Nights Searched For
Microvariability [Electronic Table]} \tablehead{ \colhead{Mrk 335}
& \colhead{UT} & \colhead{JD}
& \colhead{V (magnitude)} & \colhead{Error} & \colhead{FWHM}\\
\colhead{ } & \colhead{ } & \colhead{ } & \colhead{ } &
\colhead{$\pm$ } & \colhead{(arcsec)}}
\startdata

contact author for data

 \enddata

\end{deluxetable}
\clearpage
\begin{deluxetable}{lrrrrcrrrrrr}
\tablewidth{0pt}
\tablecaption{V-Magnitudes During Nights Searched For
Microvariability [Electronic Table]} \tablehead{ \colhead{Mrk 359}
& \colhead{UT} & \colhead{JD}   & \colhead{V (magnitude)} &
\colhead{Error} & \colhead{FWHM}\\ \colhead{ } & \colhead{ } &
\colhead{ } & \colhead{ } & \colhead{$\pm$ } & \colhead{(arcsec)}}
\startdata

contact author for data

 \enddata

\end{deluxetable}
\clearpage
\begin{deluxetable}{lrrrrcrrrrrr}
\tablewidth{0pt}
\tablecaption{V-Magnitudes During Nights Searched For
Microvariability [Electronic Table]} \tablehead{ \colhead{Mrk 478}
& \colhead{UT} & \colhead{JD}   & \colhead{V (magnitude)} &
\colhead{Error} & \colhead{FWHM}\\ \colhead{ } & \colhead{ } &
\colhead{ } & \colhead{ } & \colhead{$\pm$ } & \colhead{(arcsec)}}
\startdata

contact author for data

 \enddata

\end{deluxetable}
\clearpage
\begin{deluxetable}{lrrrrcrrrrrr}

\tablewidth{0pt}
\tablecaption{V-Magnitudes During Nights Searched For
Microvariability [Electronic Table]} \tablehead{ \colhead{Mrk 493}
& \colhead{UT} & \colhead{JD}   & \colhead{V (magnitude)} &
\colhead{Error} & \colhead{FWHM}\\ \colhead{ } & \colhead{ } &
\colhead{ } & \colhead{ } & \colhead{$\pm$ } & \colhead{(arcsec)}}
\startdata

contact author for data

\enddata

\end{deluxetable}
\clearpage
\begin{deluxetable}{lrrrrcrrrrrr}
\tablewidth{0pt}
\tablecaption{V-Magnitudes During Nights Searched For
Microvariability [Electronic Table]} \tablehead{ \colhead{NGC
4051} & \colhead{UT} & \colhead{JD}   & \colhead{V (magnitude)} &
\colhead{Error} & \colhead{FWHM}\\ \colhead{ } & \colhead{ } &
\colhead{ } & \colhead{ } & \colhead{$\pm$ } & \colhead{(arcsec)}}
\startdata

contact author for data

 \enddata

\end{deluxetable}

\clearpage

\begin{deluxetable}{lrrrrcrrrrrr}
\tablewidth{0pt}
\tablecaption{Long-Term Light Curves [Electronic Table]}
\tablehead{ \colhead{Object} & \colhead{UT}  & \colhead{JD} &
\colhead{Number of} & \colhead{V (magnitude)} & \colhead{Error} & \colhead{FWHM}\\
\colhead{ } & \colhead{ } & \colhead{ } & \colhead{Images } &
\colhead{ }& \colhead{$\pm$ } & \colhead{(arcsec)}}
 \startdata

contact author for data

\enddata

\end{deluxetable}

\clearpage
\begin{deluxetable}{llrrrc}
\tablewidth{0pt}
\tablecaption{Results of Statistical Tests\label{tbl 1-A_b}}
\tablehead{ \colhead{Name} & \colhead{Night} & \colhead{t-test} &
\colhead{Pearson}& \colhead{F-test}& \colhead{Microvariability}}
\startdata
Ark 564 &   98-Oct-19 &   97.4    &   67.4    &   60.1    &       \\
    &   98-Nov-24 &   6.20 &   8.50 &   99.3    &   ?   \\
    &   02-Oct-20    &   77.8    &   42.4    &   22.8    &       \\
Mrk 335 &   02-Oct-13    &   66.1    &   83.4    &   83.0 &       \\
    &   02-Oct14    &   28.9    &   33.7    &   66.8    &       \\
    &   02-Oct-21    &   85.9    &   77.2    &   96.4    &       \\
    &   02-Oct-22    &   27.6    &   36.3    &   71.1    &       \\
    &   02-Nov-04    &   1.30 &   5.60 &   41.7    &   ?   \\
    &   02-Nov-12    &   26.3    &   11.6    &   97.6    &       \\
Mrk 359 &   02-Nov-07    &   57.0  &   81.0  &   72.4    &       \\
    &   02-Nov-25    &   50.4    &   49.6    &   56.8    &       \\
    &   02-Dec-10    &   52.7    &   39.0  &   68.8    &       \\
    &   03-Jan-13    &   98.1    &   28.0  &   49.5    &       \\
Mrk 478 &   02-Mar-12    &   48.3    &   81.8    &   48.6    &       \\
    &   02-Apr-23    &   32.4    &   12.6    &   79.1    &       \\
    &   02-May-03    &   8.10 &   10.3    &   74.7    &       \\
    &   02-May-14    &   41.7    &   71.9    &   61.0  &       \\
    &   02-May-18    &   6.60 &   10.5    &   18.2    &       \\
    &   02-May-20    &   86.4    &   54.2    &   83.8    &       \\
    &   02-May-21    &   3.60 &   4.90 &   81.9    &   ?   \\
    &   02-May-31    &   27.7    &   43.0  &   95.1    &       \\
    &   02-Jun-01    &   89.8    &   39.0  &   86.8    &       \\
    &   02-Jun-06    &   97.3    &   86.2    &   3.10 &       \\
    &   02-Jun-09    &   92.8    &   58.9    &   94.1    &       \\
    &   02-Jun-14    &   70.8    &   47.2    &   53.9    &       \\
Mrk 493 &   02-Jun-27    &   8.30 &   7.30 &   77.2    &       \\
    &   02-Jun-28    &   47.8    &   6.60 &   82.1    &       \\
    &   03-Apr-09    &   37.6    &   11.0  &   96.9    &       \\
NGC 4051    &   03-Feb-13    &   78.3    &   77.2    &   66.8    &       \\
    &   03-Feb-20    &   0.400 &   0.300 &   16.5    &   Yes \\
    &   03-Feb-21    &   39.6    &   5.90 &   26.2    &       \\
    &   03-Mar-06    &   62.2    &   71.9    &   68.8    &       \\
    &   03-Mar-14    &   16.7    &   9.90 &   73.9    &       \\

\enddata

\end{deluxetable}

\clearpage
\begin{deluxetable}{lr}
\tablewidth{0pt}
\tablecaption{Flux Ratios Scale Factors} \tablehead{
\colhead{Object} & \colhead{Scale Factor}} \startdata

Mrk 335 &   1.14    \\
Mrk 359 &   1.93    \\
Mrk 478 &   1.09    \\
Mrk 493 &   1.38    \\
NGC 4051    &   1.55    \\
Ark 564     &   1.24    \\

\enddata

\end{deluxetable}

\clearpage
\begin{deluxetable}{lrrrrrrr}

\tablewidth{0pt}
\tablecaption{Root-Mean-Square Seasonal Variability of NLS1s}

\tablehead{\colhead{ } & \colhead{NGC} & \colhead{NGC} &
\colhead{Ark} & \colhead{Mrk} & \colhead{Mrk} & \colhead{Mrk}&
\colhead{Mrk}\\

\colhead{ } & \colhead{4051} & \colhead{4051} & \colhead{564} &
\colhead{335} & \colhead{359} & \colhead{478}& \colhead{493}\\}

\startdata

    &   0.036   &   0.014   &   0.025   &   0.023   &   0.026   &   0.025   &   0.050   \\
    &   0.045   &       &   0.054   &       &       &       &      \\
    &   0.048   &       &   0.147   &       &       &       &       \\
    &       &       &   0.045   &       &       &       &       \\
    &       &       &   0.014   &       &       &       &        \\
    &       &       &       &       &       &       &           \\
Ave (scaled)    &   0.043   &   0.022   &   0.071   &   0.026   &   0.050   &   0.027   &   0.069    \\

\enddata

\end{deluxetable}

\clearpage
\begin{deluxetable}{lrrrrrrrrr}

\tablewidth{0pt}
\tablecaption{Root-Mean-Square Seasonal Variability of BLS1s}
\tablehead{\colhead{ } & \colhead{NGC} & \colhead{NGC} &
\colhead{NGC} & \colhead{NGC} & \colhead{Mrk} & \colhead{Mrk}&
\colhead{3C390.3} & \colhead{3C 273 } &
\colhead{Average}\\\colhead{ } & \colhead{3783} & \colhead{4151} &
\colhead{5548} & \colhead{7469} & \colhead{ 279} & \colhead{509}&
\colhead{ } & \colhead{ } & \colhead{}\\\colhead{ } &
\colhead{IAW} & \colhead{IAW} & \colhead{IAW} & \colhead{IAW}&
\colhead{IAW} & \colhead{IAW} &
\colhead{see ref}& \colhead{see ref} & \colhead{ }\\
\colhead{Ref*} & \colhead{a} & \colhead{b} & \colhead{c} &
\colhead{d}& \colhead{e} & \colhead{f} & \colhead{g}& \colhead{h}
& \colhead{ }}
 \startdata

    &   0.085   &   0.032   &   0.137   &   0.026   &   0.087   &   0.132   &   0.061   &   0.030   &       \\
    &       &   0.080    &   0.136   &       &   0.079   &   0.076   &   0.151   &   0.016   &       \\
    &       &   0.090    &   0.099   &       &       &   0.142   &   0.107   &   0.059   &       \\
    &       &   0.041   &   0.172   &       &       &   0.087   &       &   0.073   &       \\
    &       &   0.106   &   0.100 &       &       &   0.086   &       &   0.026   &       \\
    &       &   0.037   &   0.112   &       &       &       &       &   0.082   &       \\
    &       &       &   0.085   &       &       &       &       &   0.040   &       \\
    &       &       &   0.161   &       &       &       &       &   0.067   &       \\
    &       &       &   0.113   &       &       &       &       &   0.014   &       \\
    &       &       &   0.107   &       &       &       &       &   0.023   &       \\
    &       &       &   0.165   &       &       &       &       &   0.032   &       \\
    &       &       &   0.173   &       &       &       &       &   0.077   &       \\
    &       &       &   0.130   &       &       &       &       &       &       \\
    &       &       &       &       &       &       &       &       &       \\
Ave     &   0.085   &   0.064   &   0.130   &   0.026   &   0.083   &   0.104   &   0.106   &   0.045   &   0.078   \\

\enddata

\end{deluxetable}

\clearpage

\begin{figure}
\plotone{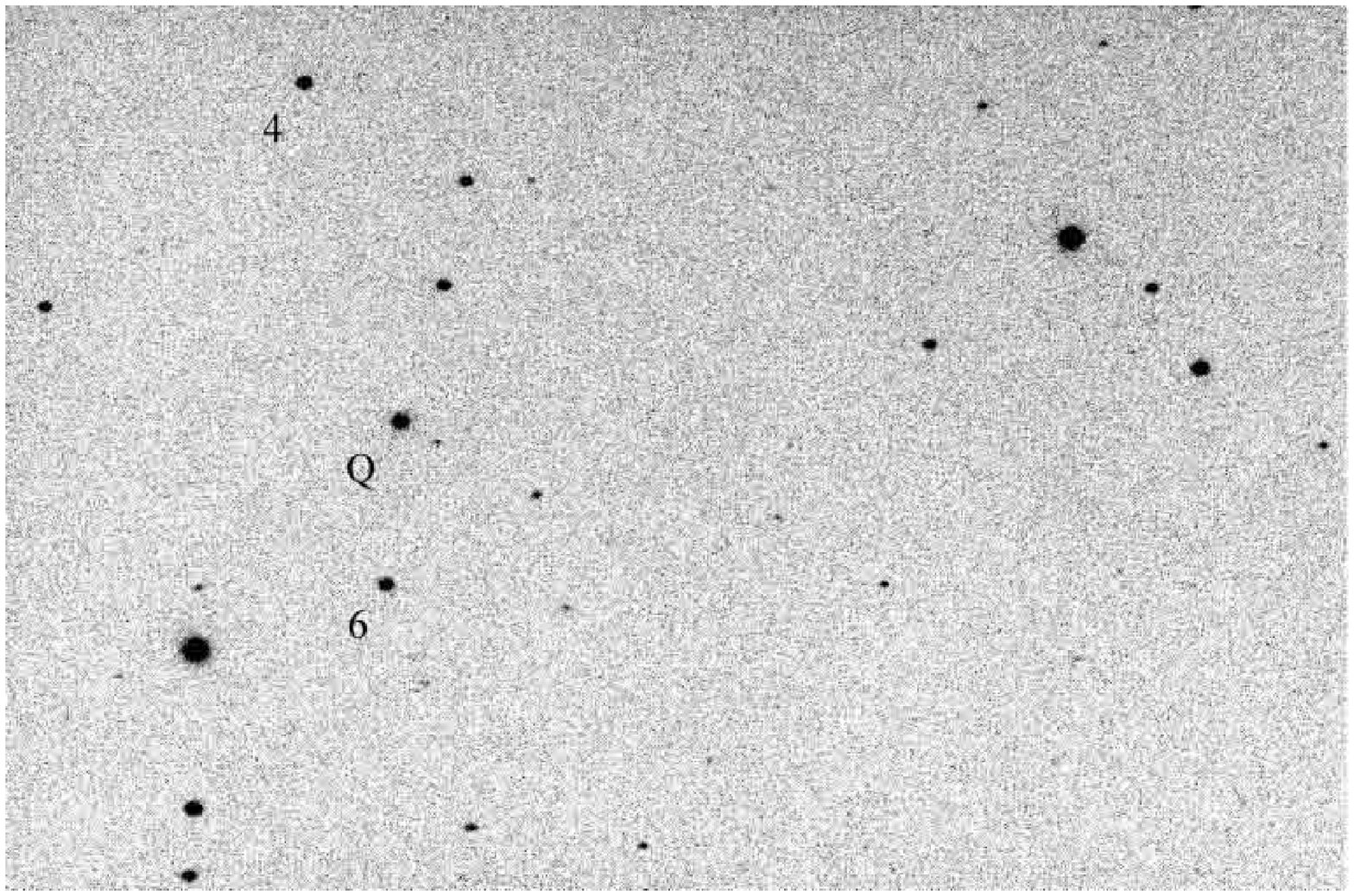} \caption{Comparison stars for Mrk 335}
\end{figure}
\begin{figure}
\plotone{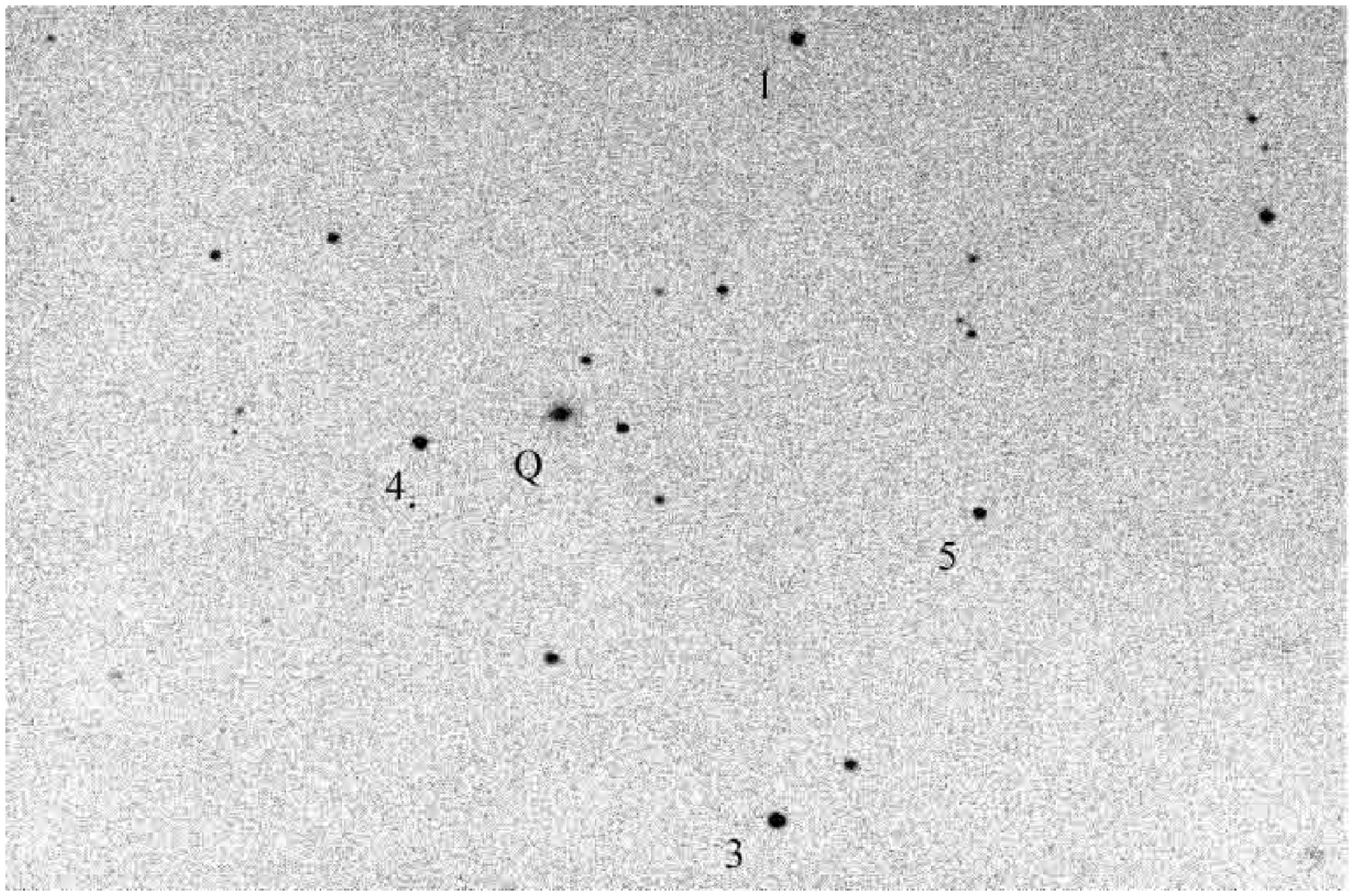} \caption{Comparison stars for Mrk 359}
\end{figure}
\begin{figure}
\plotone{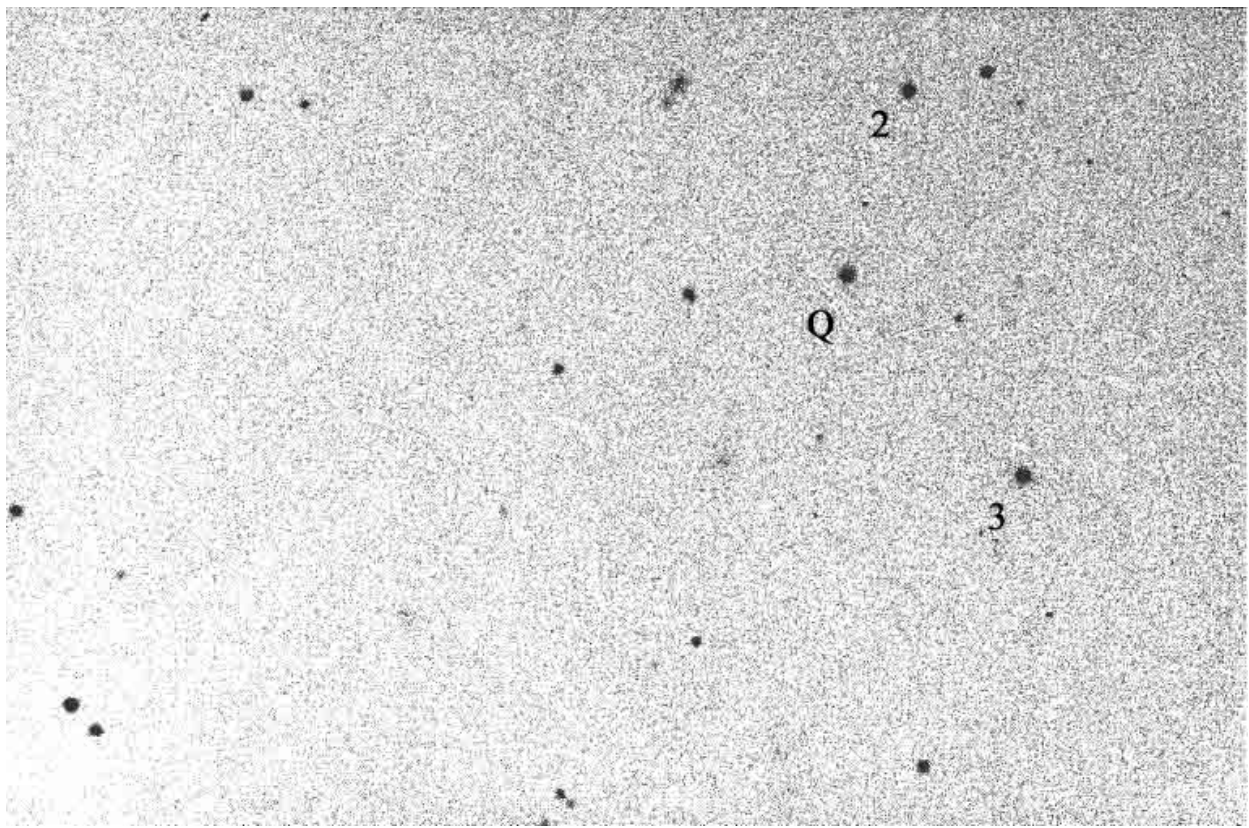} \caption{Comparison stars for Mrk 478}
\end{figure}
\begin{figure}
\plotone{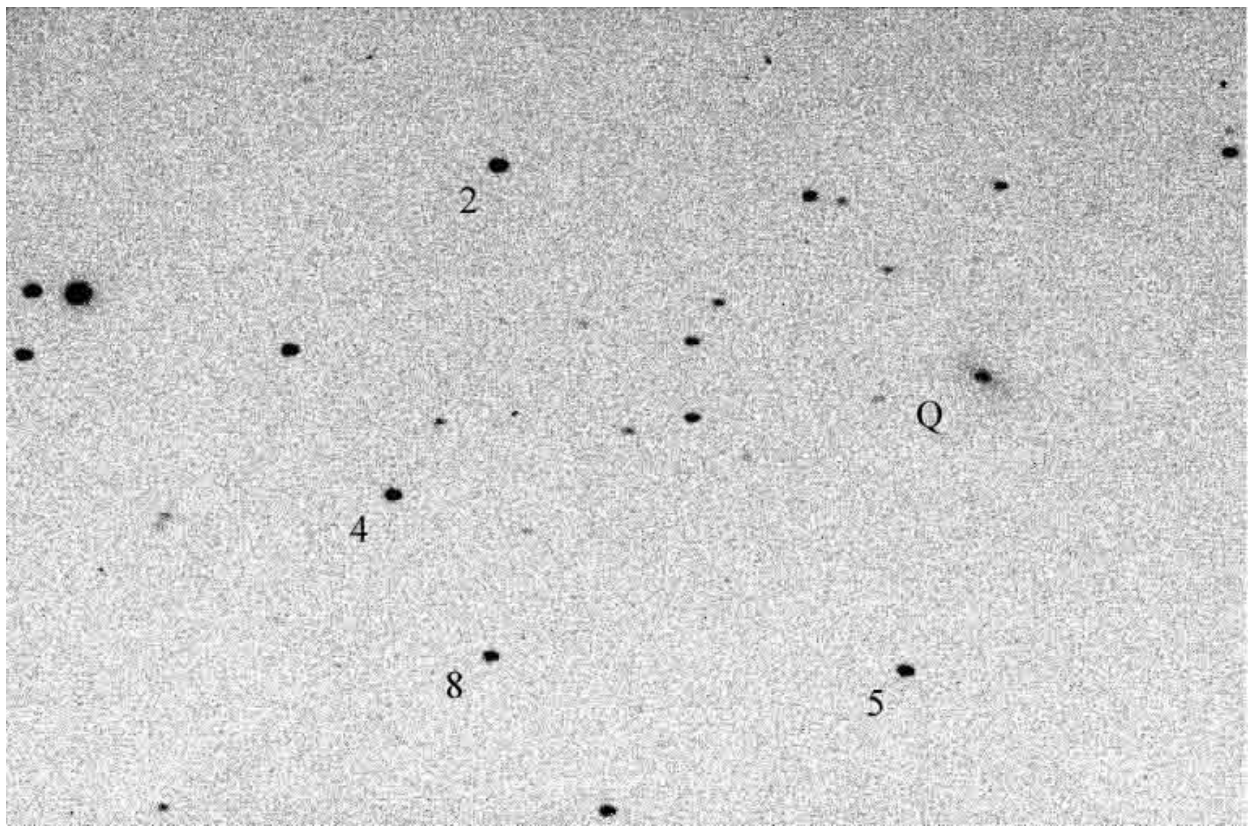} \caption{Comparison stars for Mrk 493}
\end{figure}
\begin{figure}
\plotone{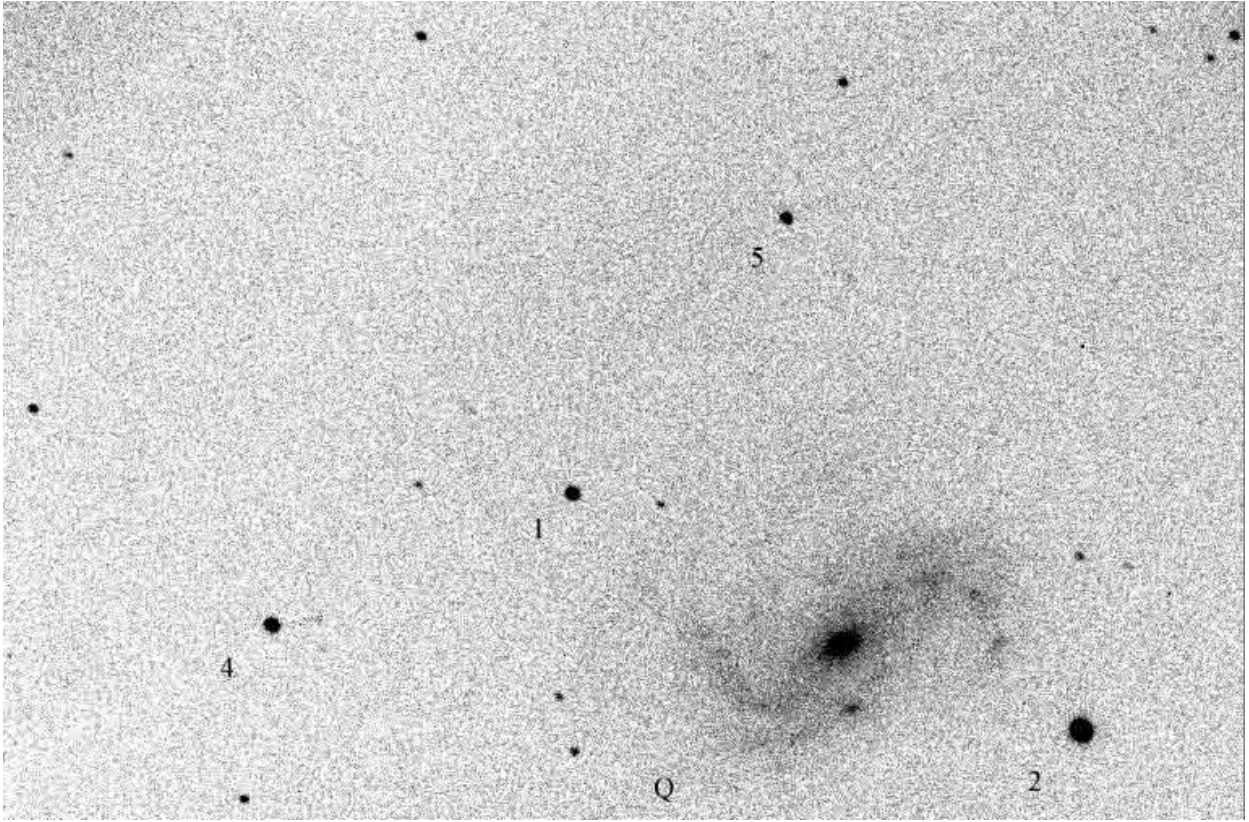} \caption{Comparison stars for NGC 4051}
\end{figure}
\begin{figure}
\plotone{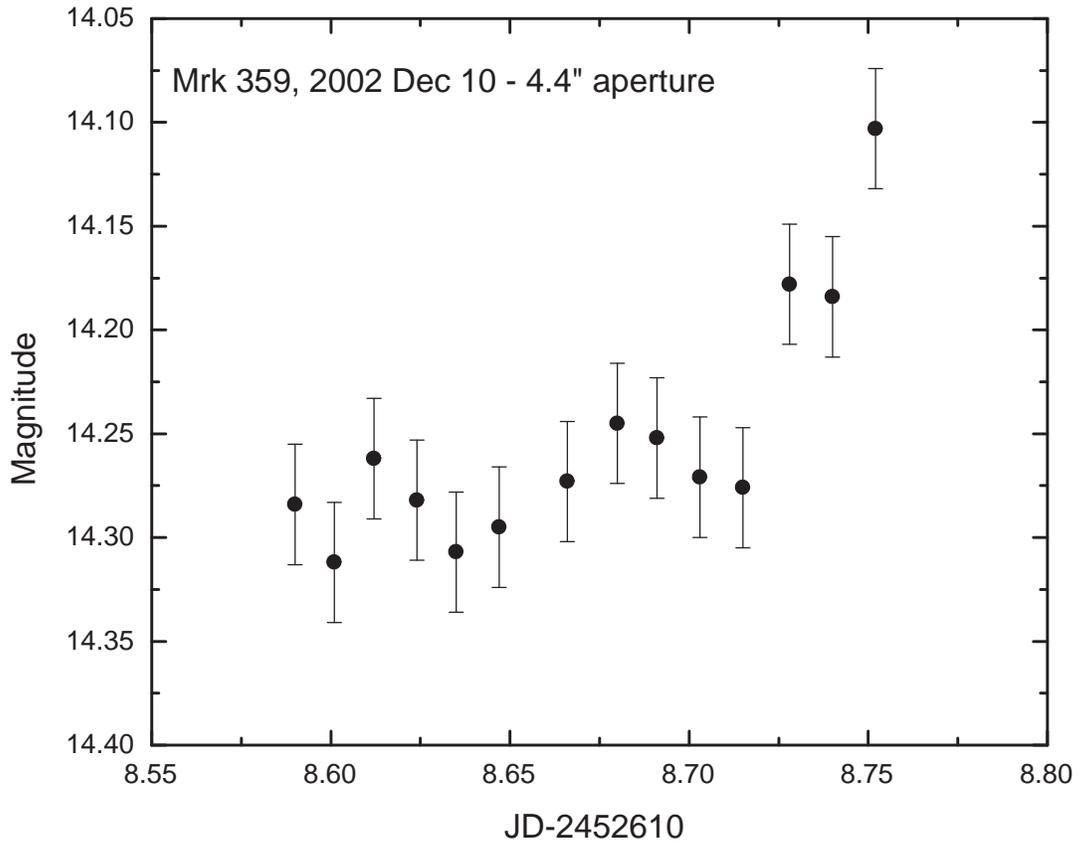} \caption{Apparent microvariability of Mrk 359
when measured with a 4.4 arcsecond aperture on 2002 December 10.}
\end{figure}
\begin{figure}
\plotone{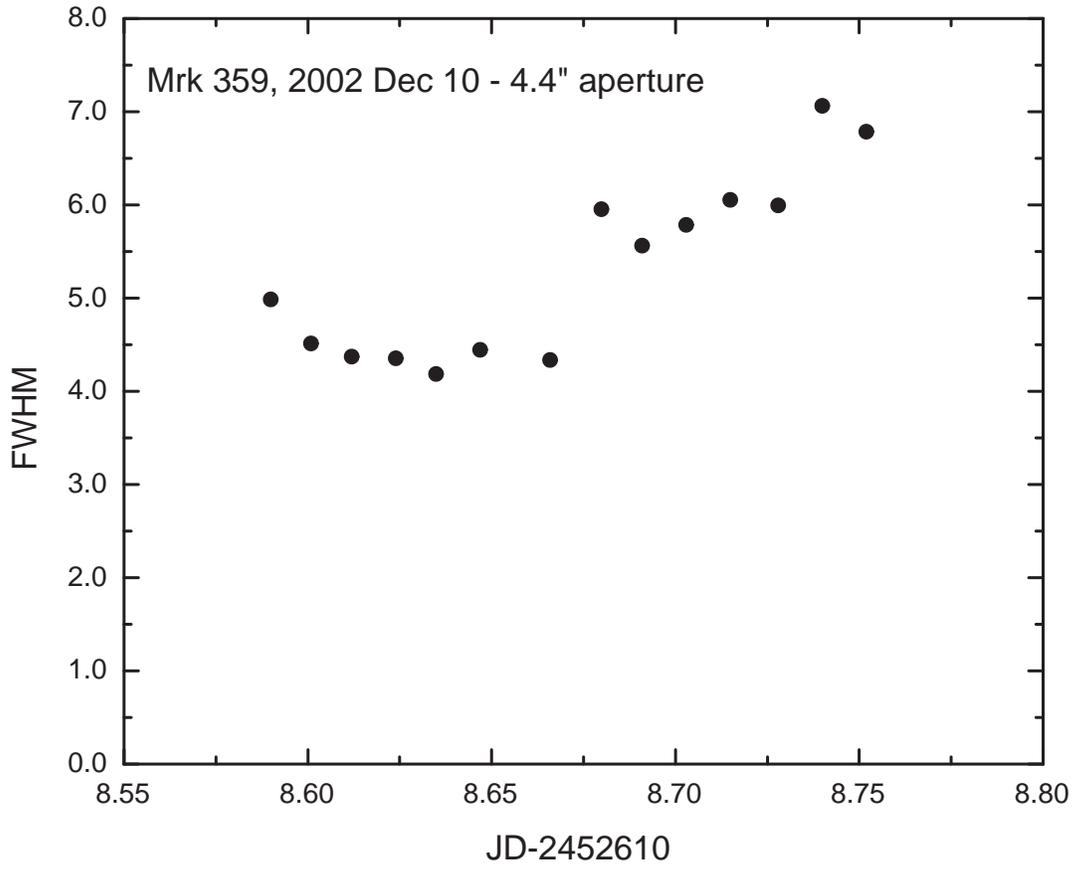} \caption{Variation of image quality for Mrk
359 on 2002 December 10.} \end{figure}
\begin{figure}
\plotone{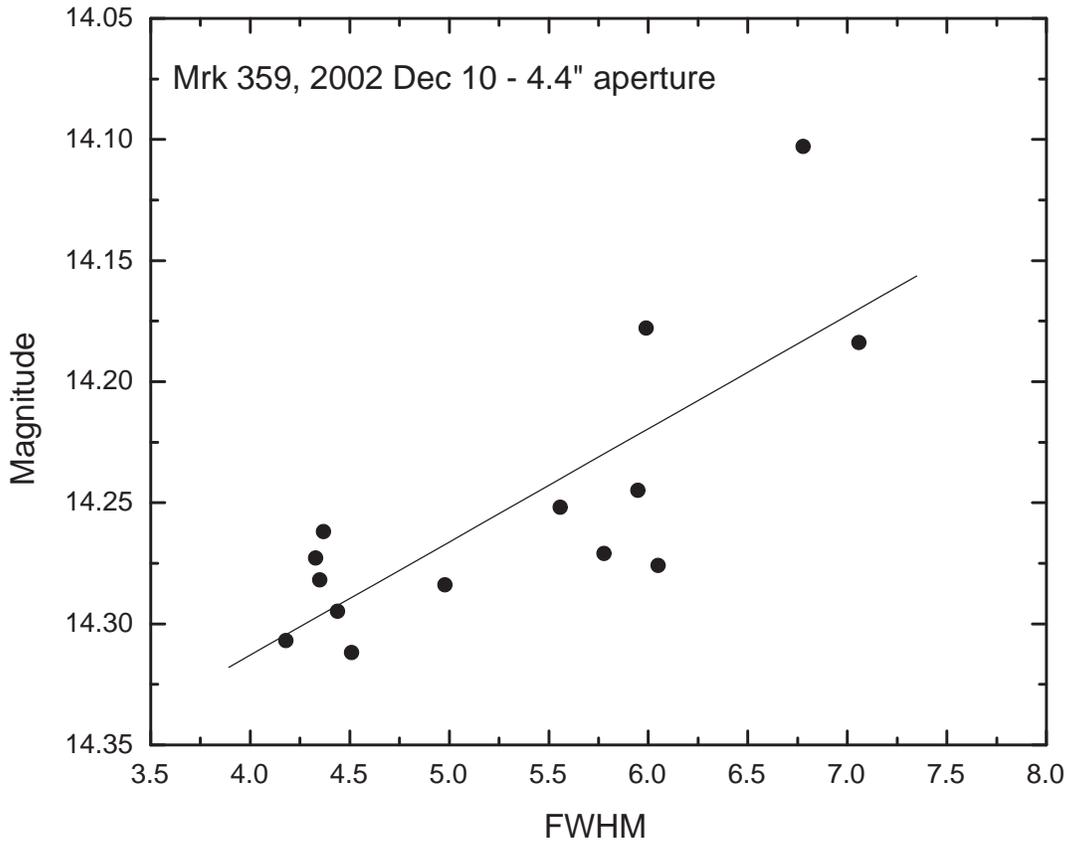} \caption{Variation of measured magnitude with
image quality for Mrk 359 when measured with a 4.4 arcsecond
aperture.} \end{figure}
\begin{figure}
\plotone{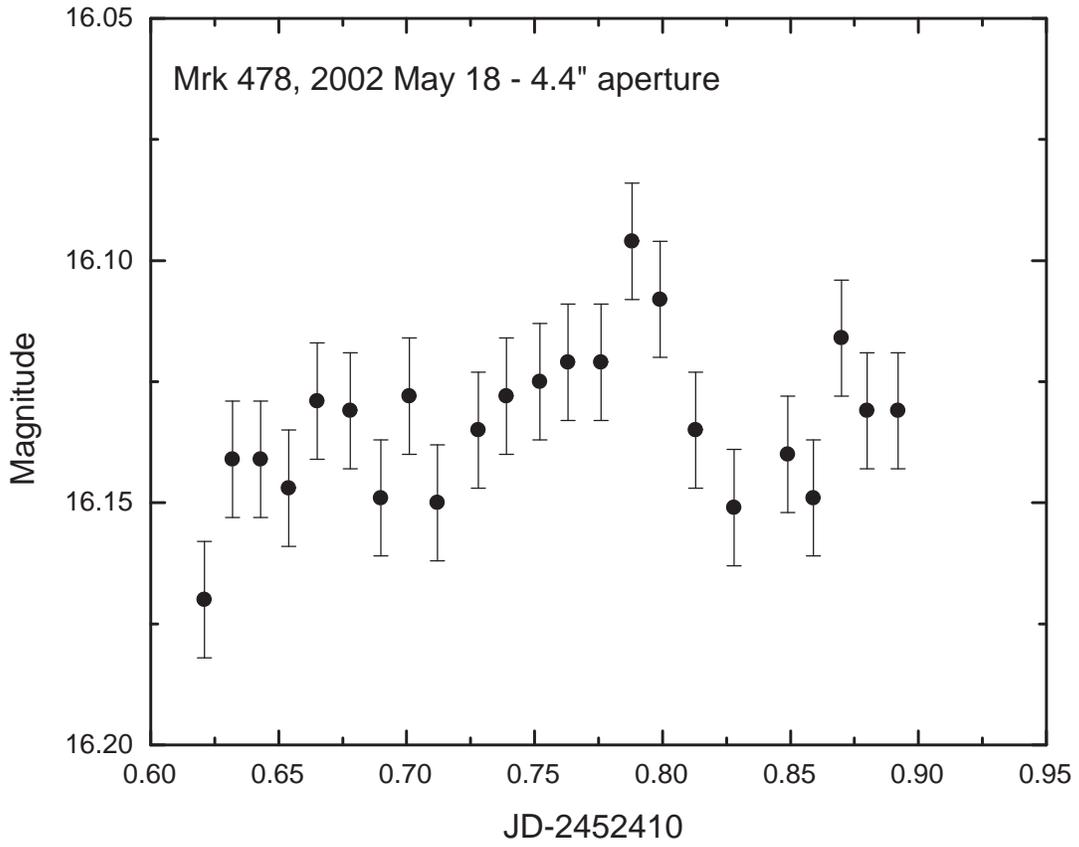} \caption{Apparent microvariability of Mrk 478
on 2002 May 18 when measured with a 4.4 arcsecond aperture.}
\end{figure}
\begin{figure}
\plotone{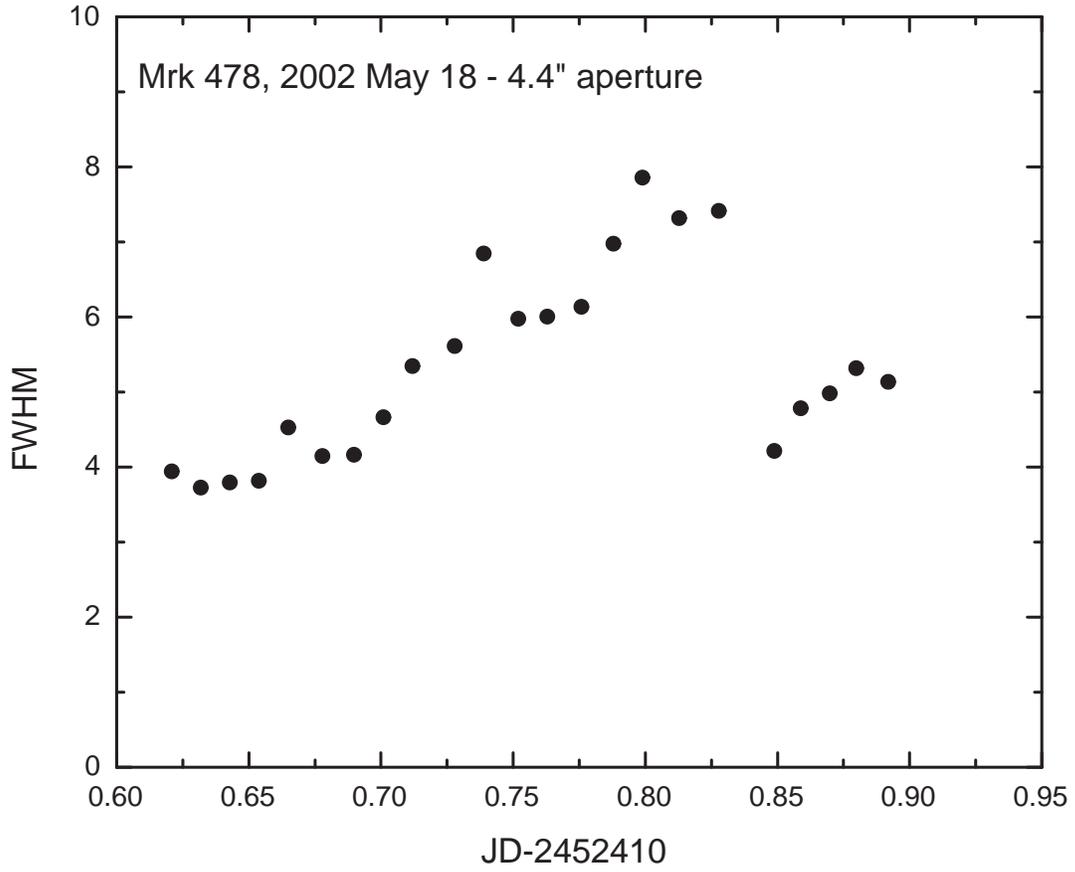} \caption{Variation of measured magnitude with
image quality for Mrk 478 on 2002 May 18.} \end{figure}
\begin{figure}
\plotone{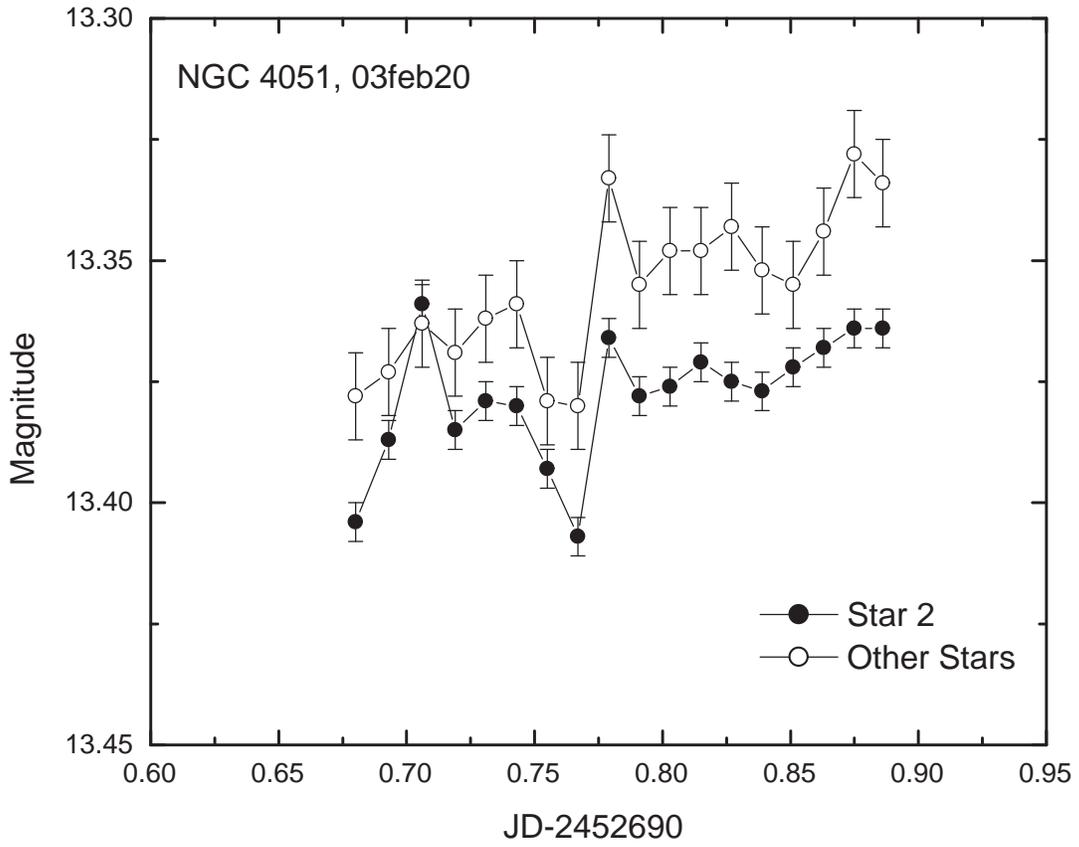} \caption{V-band light curve for NGC 4051 on
2003 February 20.  The solid circles are the magnitudes measured
relative to just star 2; the open circles are the magnitudes
relative to the other comparison stars.}
\end{figure}
\begin{figure}
\plotone{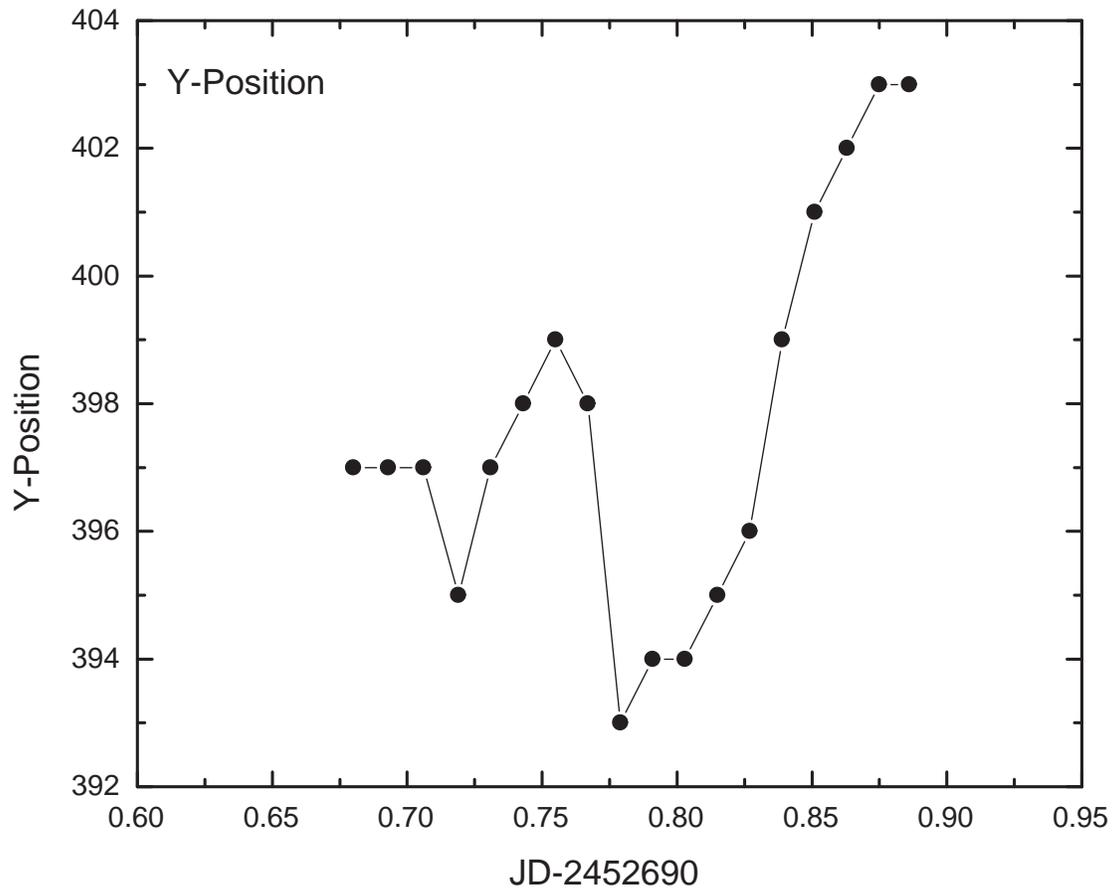} \caption{Variation of the y-position of the
nucleus of NGC 4051 on the chip during the night of 2003 February
20.}
\end{figure}
\begin{figure}
\plotone{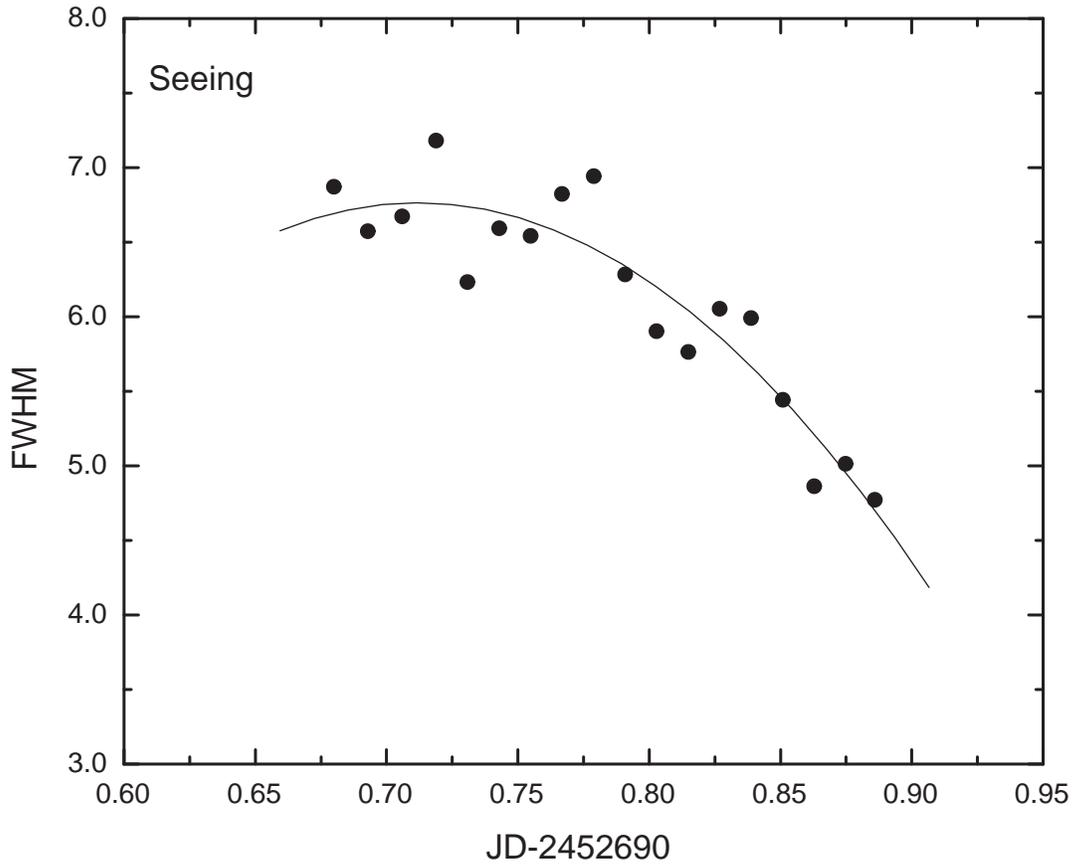} \caption{Variation of the image quality for
images of NGC 4051 during the night of 2003 February 20.}
\end{figure}
\begin{figure}
\plotone{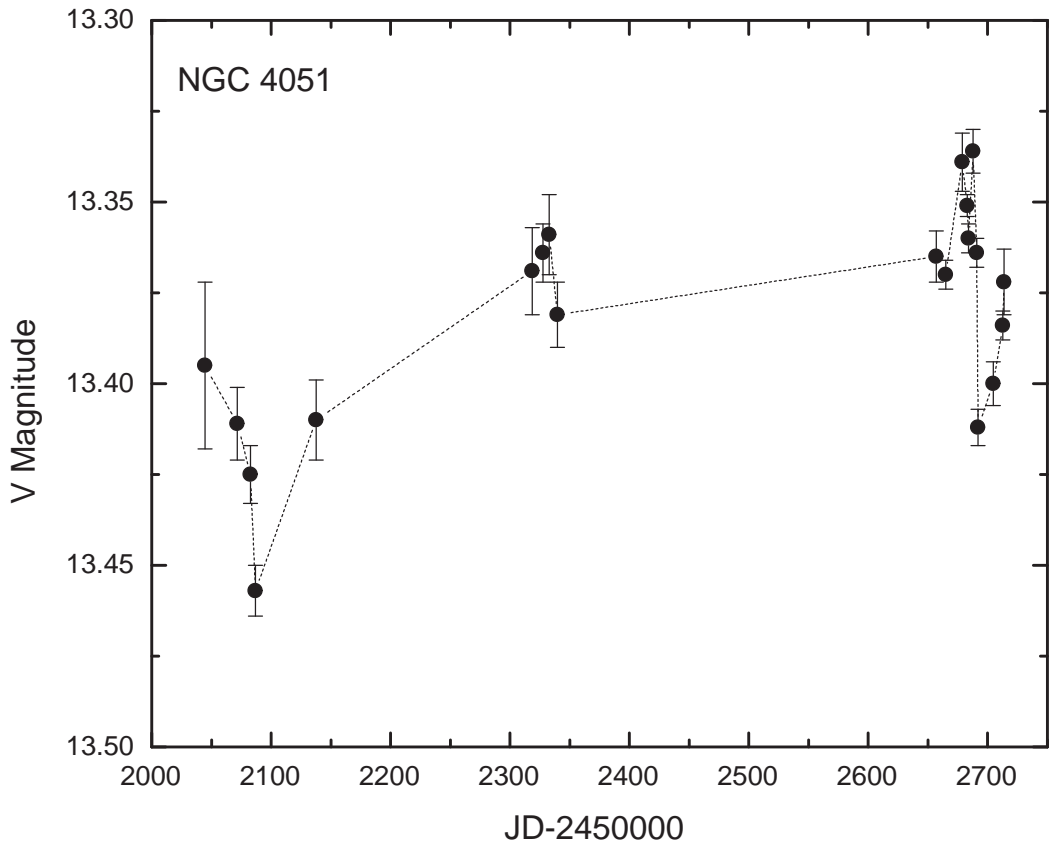} \caption{Long-term light curve for NGC 4051.}
\end{figure}
\begin{figure}
\plotone{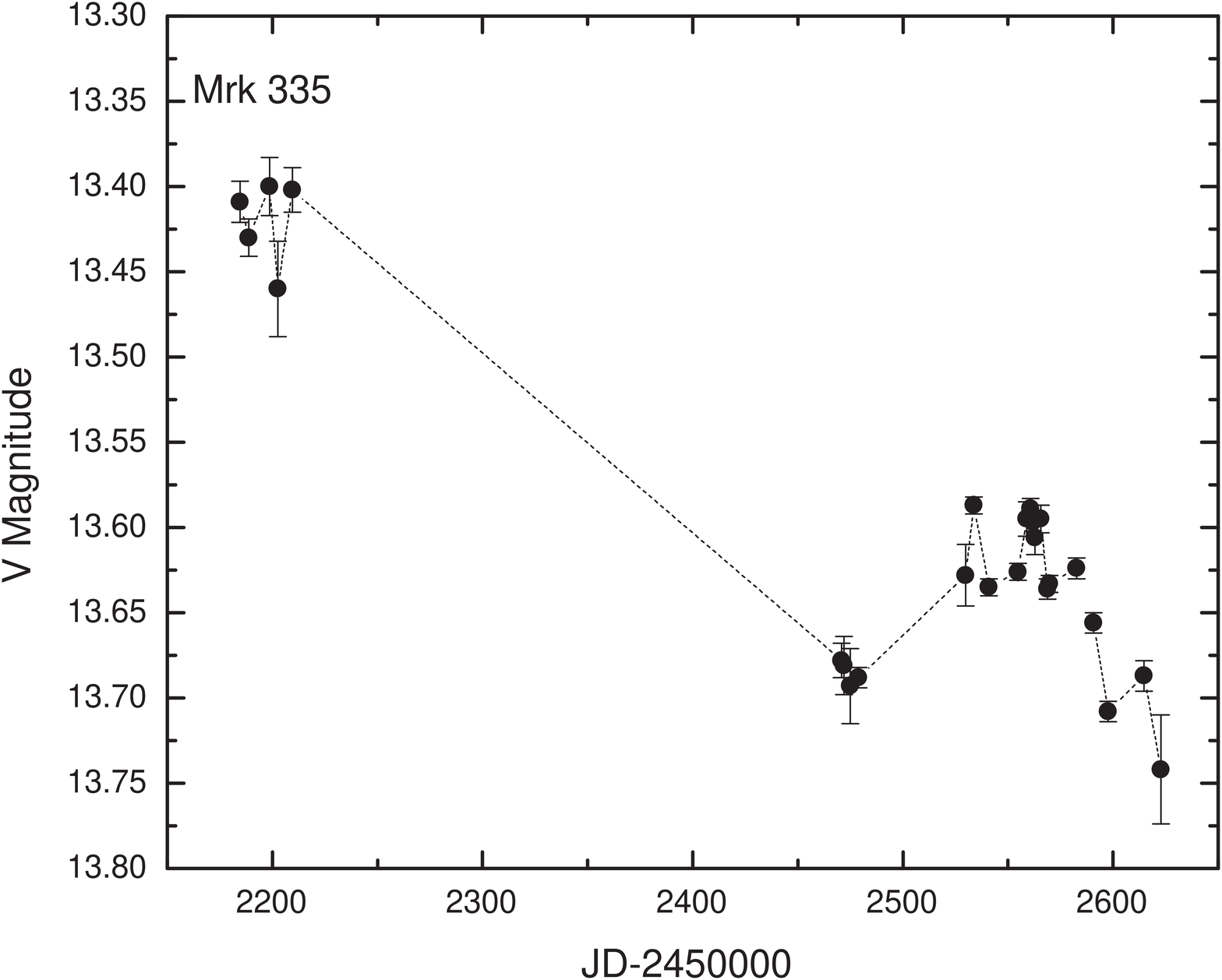} \caption{Long-term light curve for Mrk 335.}
\end{figure}
\begin{figure}
\plotone{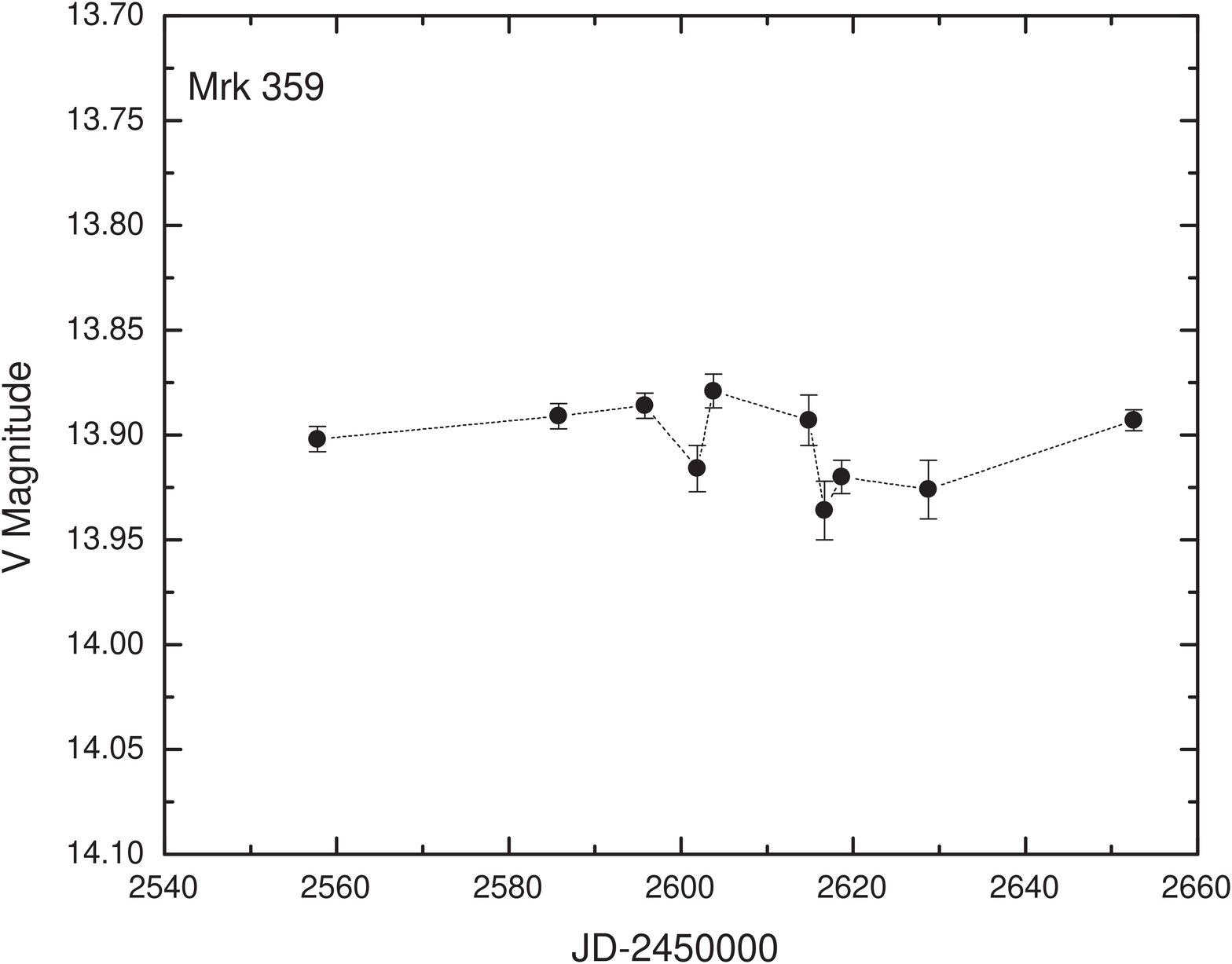} \caption{Long-term light curve for Mrk 359.}
\end{figure}
\begin{figure}
\plotone{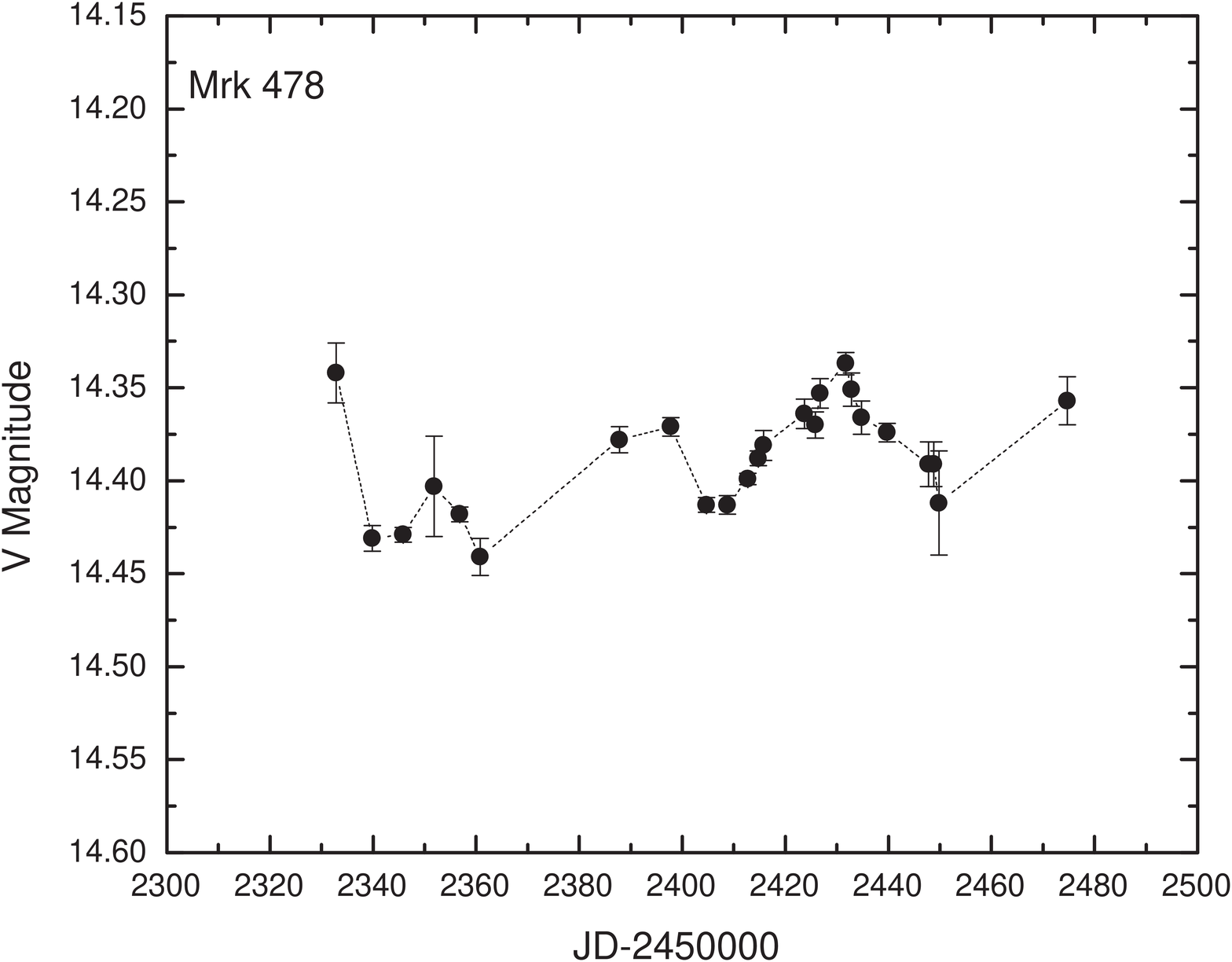} \caption{Long-term light curve for Mrk 478. }
\end{figure}
\begin{figure}
\plotone{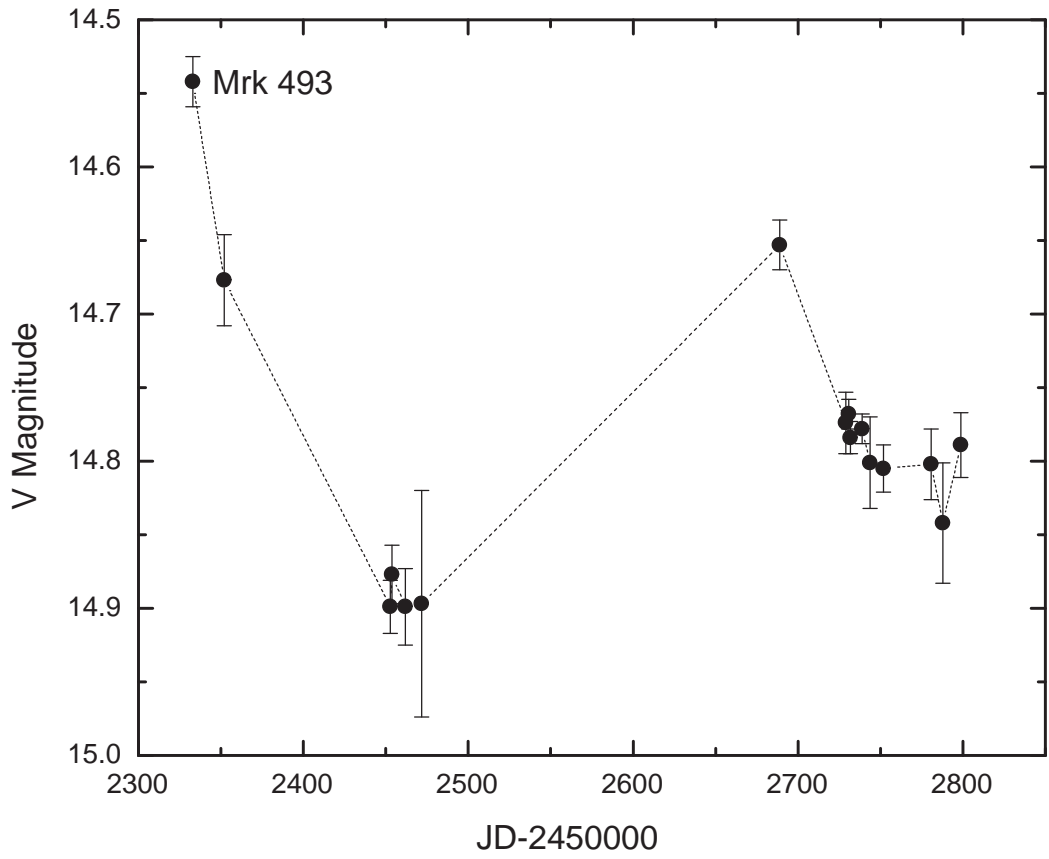} \caption{Long-term light curve for Mrk 493.}
\end{figure} \clearpage
\begin{figure}
\epsscale{.85} \plotone{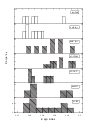} \caption{Distributions of
root-mean-square V-band variability amplitudes for NLS1s (open
squares) and BLS1s (hashed squares). For individual objects the
seasonal variability is shown.  For the NLS1 average and BLS1
average each square represents the mean of the variability over
the seasons given in Tables 10 and 11.}
\end{figure}

\clearpage

Fig. 19.--- Distributions of root-mean-square V-band variability
amplitudes for NLS1s (open squares) and BLS1s (hashed squares).
For individual objects the seasonal variability is shown.  For the
NLS1 average and BLS1 average each square represents the mean of
the variability over the seasons given in Tables 10 and 11.

\end{document}